\documentstyle{mn}
\title[SuperCOSMOS Sky Survey I: Introduction \& Description]{The 
SuperCOSMOS Sky Survey. Paper I: \\ Introduction and Description}
\author[N.C.\ Hambly et al.]{N.C.\ Hambly$^{\rm 1}$, H.T.\ MacGillivray$^{\rm 1}$,
M.A.\ Read$^{\rm 1}$, S.B.\ Tritton$^{\rm 1}$, E.B.\ Thomson$^{\rm 1}$,\cr
B.D.\ Kelly$^{\rm 2}$, D.H.\ Morgan$^{\rm 3}$
R.E.\ Smith$^{\rm 3}$, S.P.\ Driver$^{\rm 4}$, J.\ Williamson$^{\rm 1}$,\cr
Q.A.\ Parker$^{\rm 1}$, M.R.S.\ Hawkins$^{\rm 1}$,
P.M.\ Williams$^{\rm 1}$, A.\ Lawrence$^{\rm 3}$\\
$^1$Wide Field Astronomy Unit, Institute for Astronomy, University of Edinburgh, Blackford Hill, Edinburgh, EH9~3HJ\\
$^2$UK ATC, Royal Observatory, Blackford Hill, Edinburgh, EH9~3HJ\\
$^3$Institute for Astronomy, University of Edinburgh, Blackford Hill, Edinburgh, EH9~3HJ\\
$^4$School of Physics and Astronomy, University of St.\ Andrews, North Haugh,
St.\ Andrews, Fife, KY16~9SS
}

\date{Accepted ---. 
      Received ---;
      in original form ---}

\begin{document}

\maketitle

\begin{abstract}
In this, the first in a series of three papers concerning the SuperCOSMOS
Sky Survey (SSS), 
we give an introduction and user guide to the survey programme.
We briefly describe other wide--field surveys and compare with our own.
We give
examples of the data, and make a comparison of the accuracies of the
various image parameters available with those from the other surveys
providing similar data; we show that the SSS database and interface offer 
advantages over these surveys.
Some science applications of the data are also described and some
limitations discussed. The series
of three papers constitutes a comprehensive description and user guide
for the SSS.
\end{abstract}

\section{Introduction}
\label{intro}

This is the first paper in a series of three concerning the SuperCOSMOS Sky
Survey (hereafter SSS) programme\footnote{database 
available online at \verb+http://www-wfau.roe.ac.uk/sss+}. 
This ambitious project aims to digitise the
sky in three colours~(BRI), one colour~(R) at two epochs, via automatic scans
of sky atlas photographs. The source photographic material comes from Schmidt
telescopes, observations being taken during the second half of the twentieth
century (see later). Ultimately we aim to digitise the entire sky, but due to
the time required to do so data are being `released' to the community as they
become available. The first release of data consisted of 5000~square degrees
of the southern sky at high Galactic latitude ($|b|>60^{\circ}$). This 
became known
as the South Galactic Cap (hereafter SGC) survey. All three papers in this
series make reference to data from the SGC survey, but are generally relevant
to the SSS as a whole. This paper (Paper~{\sc I}) is intended as a general
introduction and user guide to the SSS. Paper~{\sc II} (Hambly, Irwin \&
MacGillivray~2001a) describes image detection, parameterisation, classification
and photometry while Paper~{\sc III} (Hambly et al.~2001b) is concerned with
astrometry. These latter two papers contain much technical information about the
SSS. This paper summarises that information and compares with similar data from
other survey programmes. We also describe how to access the data and give 
examples of results from the SSS.

Astronomical photography, which revolutionised astronomy in the late
nineteenth and early twentieth centuries, has a history almost as long as that
of photography itself (Lankford~1984 and references therein). 
Digitised photographic sky surveys revolutionised astronomy in the late 
twentieth century; their evolution over the last few decades has been dictated 
by developments in microdensitometry and digital electronics 
(e.g.\ Klingsmith~1984; MacGillivray \& Thomson~1992). Of course, in
the modern era the sky is being digitised directly, bypassing photography
altogether. This brings many advantages, not the least of which are the ability
to survey faster, deeper and at wavelengths other than the optical -- e.g.\
the Sloan Digital Sky Survey (SDSS, York et al.~2000) and the Two--Micron
All--Sky Survey (2MASS, Kleinmann et al.~1994). However, it must be emphasised
that for time--dependent phenomena for example (e.g.~object position and
brightness) photographic sky atlases represent a vast archive of invaluable
measurements for {\em billions} of objects. This has been suitably demonstrated
recently by the addition of Astrographic Catalogue measurements of the Carte
du Ciel plates to Tycho positions from the Hipparcos mission
(Urban, Corbin \& Wycoff~1998; H\o g et al.~2000
and references therein). This juxtaposition of
century--old photographic astrometry with a state--of--the--art space--based
mission has yielded stellar proper motions on an unprecedented scale: 
{\em millions} of objects have proper motions measured to a precision of
2.5~mas~yr$^{-1}$.

It was the development of the Schmidt telescope in the 1930s that revolutionised
wide--angle photographic surveys (e.g.~Cannon~1995). Along with developments
in fast photographic emulsions, these enabled sky atlas production to faint
limiting magnitudes. The 1.2m Palomar Oschin, 1.0m ESO and 1.2m UK Schmidt
Telescopes have been systematically photographing the whole sky
over the last half--century. Photographs have been taken in the blue, red and
near--infrared passbands, with both blue and red passbands having observations
at two epochs per field. For a concise summary of these photographic
survey programmes, see Morgan~(1995). Given the vast numbers 
($>10,000$) of plates in these
sky atlases, it is clear that it is only through digitisation that these surveys
can be fully exploited.

There are several major digitisation programmes completed or currently in
progress. These projects consist of scanning various subsets of the available
sky survey atlas photographs (in some cases glass or film copies of the original
Schmidt photographs) and release of various data to the community.
The scope of the various projects and their emphasis is
different in each case. We now consider each in turn:

\subsection{APM (Cambridge, UK)}
\label{apm}

The Automatic Plate Measuring (APM) machine is a `flying--spot' microdensitometer
with pixel size of $7.5\mu$m (or 0.5~arcsec at the Schmidt plate scale of
$\sim67$~arcsec~mm$^{-1}$). For details concerning the APM hardware, see
Kibblewhite et al.~(1984); more information is also available at
\verb+http://www.ast.cam.ac.uk/~mike/casu/apm/apm.html+. The APM facility has
produced the APM Northern Sky Catalogue (Irwin \& McMahon~1992) which is based
on scans of the first epoch Palomar (POSS--I)~O and~E glass copy atlases and
covers the northern sky down to Galactic latitudes $|b|\geq20^{\circ}$. This
programme is currently being extended into the southern hemisphere using the
SERC--J/EJ and SERC--ER/AAO--R surveys. The APM sky catalogues provide paired
lists of blue and red objects with morphological parameters and image
classification. Currently, they do not include pixel data, object proper motions
or any information in the near--infrared. For more details, see the web pages
accessed via the above URL.

\subsection{APS (Minneapolis, USA)}
\label{aps}

The Automated Plate Scanner (APS) machine is described by Pennington et 
al.~(1993) and is similar in design to the APM. This machine is also undertaking
a scanning programme based on the POSS--I glass copies, but supplemented by
the corresponding second epoch red plates obtained for the Luyten proper motion
surveys (e.g.~Luyten~1979 and references therein) and ultimately POSS--II.
All 664 POSS--I fields having Galactic latitude $|b|>20^{\circ}$ have been
included. The APS object catalogue database (see \verb+http://aps.umn.edu/+) is
similar in content to that of the APM; proper motions will additionally be 
available in the future. Also, restricted numbers of plates have thresholded
pixel data available online (the pixel size of the APS scans is 0.8~arcsec).
These thresholded images have all sky areas set to zero intensity. Once again,
no near--infrared data are available. For more information concerning the APS
database, see Cornuelle et al.~(1997).

\subsection{COSMOS (Edinburgh, UK)}
\label{cos}

The COSMOS machine (e.g.~MacGillivray \& Stobie~1984), also a `flying--spot'
microdensitometer, was a development of the original GALAXY plate
measuring machines (e.g.\ Murray \& Nicholson~1975).
COSMOS ceased operation in late
1993 and has now been superseded by SuperCOSMOS (see Section~\ref{scos}). The
older machine had pixel size 1.1~arcsec and undertook a single 
colour~(B$_{\rm J}$) survey of the southern sky using the SERC--J/EJ atlas glass
copies and UK Schmidt `short' red (SR) originals at low Galactic latitude.
The resulting database, the COSMOS/UKST Catalogue of the Southern Sky,
included all fields at or south of the equator and having Galactic latitude
$|b|>10^{\circ}$ (J) or $|b|<10^{\circ}$ (SR). For more details, see Yentis et
al.~(1992) and also Drinkwater, Barnes \& Ellison~(1995); an online database
is available at \verb+http://xip.nrl.navy.mil/+. The
object catalogue contains position, morphological, photometric and classification
information for each object in one colour (B$_{\rm J}$ or~R depending on Galactic
latitude). No pixel data, proper motions or near--infrared data are available.

\subsection{DSS (Baltimore, USA)}
\label{dss}

Probably the greatest in scope of all the digitisation programmes is the
Digitised Sky Survey at the Space Telescope Science Institute (Lasker~1992). This
project was originally set up to provide guide stars for use with the Hubble
Space Telescope (Jenker, Russell \& Lasker~1988). Two modified PDS machines
(known as Guide Star Automatic Measuring MAchines, or GAMMAs)
were constructed for the purpose of scanning the photographic material, 
and the first generation
scanning programme was implemented with 1.7~arcsec pixels. The surveys scanned
consisted of the SERC--J/EJ, the POSS--I~E and short exposure plates for fields
having low Galactic latitude. The original Guide Star Catalogue (GSC--I, 
e.g.~Lasker et al.~1990; Russell et al.~1990; Jenker et al.~1990) has since
proven to be a major astronomical resource, far beyond its original goal of
supporting HST operations. In addition, it was quickly realised that there was
a large demand for the pixel data from the GSC--I scans, and the DSS has also
become a major astronomical resource (for more information and online database
access, see \verb+http://archive.stsci.edu/dss/+). The volume of raw pixel
data in the DSS--I necessitated
the use of image compression techniques in order
to make distribution possible (see Section~\ref{hcomp}). 
The second generation survey
(DSS--II) is now nearly complete. This includes modified scanning hardware (the
major change from DSS--I is that the pixel size is now 1~arcsec) and the addition
of multiple colours and epochs from the POSS--II material in the northern
hemisphere, and SERC--I/SERC--ER/AAO--R material in the south. A number of survey
databases are in production. The second generation Guide Star Catalogue
(GSC--II) is under construction (McLean et al.~1997) and will include~BRI
colours as well as proper motion information, for all objects down to m~$\sim18$.
This catalogue is envisaged as being important for supporting various ground--
and space--based instruments (e.g.~Lasker et al.~1995). Compressed pixel data in
the R band are being made available online at the above URL to supplement that
available from DSS--I. Moreover, the Palomar--STScI Digital Sky Survey (DPOSS)
is a project to archive and analyse the raw pixel data in each of the three
colours (BRI) to the plate limits in every POSS--II survey field (e.g.~Weir~1995
and references therein; Djorgovski et al.~1997). Once complete,
the three--colour northern sky catalogue is expected to be released to the
astronomical community.

\subsection{PMM (Flagstaff, USA)}
\label{pmm}

The Precision Measuring Machine (PMM) programme is operated at the United States
Naval Observatory (USNO). The PMM uses CCD detectors to image (with 0.9~arcsec
pixels) large regions of pairs of photographs simultaneously and 
very quickly; the machine and 
programme have been designed primarily with astrometric applications in mind.
In addition to the CD--ROM stellar object catalogue products (containing
paired lists of objects between blue and red plates with position and colour
information only) -- e.g.~USNO--A1.0 (Canzian~1997), now superseded by 
USNO--A2.0 (Monet~1998) -- the project now provides `near--line'
access to all the raw pixel data from the plate scans
(see \verb+http://www.nofs.navy.mil/data/FchPix/cfra.html+).
Future products will
include updated CD--ROM sets with object lists containing galaxies as well as
stars, morphological information and proper motion data (e.g.\ USNO--B)

\subsection{SuperCOSMOS (Edinburgh, UK)}
\label{scos}

The SuperCOSMOS machine is operated at Edinburgh and is the successor to COSMOS.
SuperCOSMOS is a fast, high precision plate scanning facility with 0.67~arcsec
pixels, 15--bit digitisation and accurate positional capability (e.g.~see Miller
et al.~1992 and Hambly et al.~1998). The origins of the survey programme (SSS) 
can be traced back several years (e.g.~Hawkins~1992), and the SSS is 
the subject 
of this series of papers. Online database access and much additional information
is available at \verb+http://www-wfau.roe.ac.uk/sss+. In setting up the SSS
programme, our intention has been to combine the best features of all of the
above programmes to produce a versatile and easily used survey product with
object parameter accuracy limited, as far as possible, by the source photographic
material.

\subsection*{ }

The above provides a very brief introduction to the enormous subject of
digitised photographic sky surveys. For much more information, the reader is
referred to conference proceedings by Capaccioli~(1984), MacGillivray \& 
Thomson~(1992), MacGillivray et al.~(1994), Chapman et al.~(1995) and McLean et
al.~(1997a). For a general introduction to the subject of digitised sky surveys
from Schmidt plates in particular, see Lasker~(1995) and references therein.

\section{Description}
\label{methods}

\subsection{Summary of photographic material}

For a complete description of the UK/ESO/Palomar Schmidt photographic atlases,
see Morgan~(1995) and references therein. In Table~\ref{atlases} we show the
surveys which currently comprise the southern hemisphere SSS; in addition we
show the corresponding northern hemisphere surveys that may be scanned. Note
that, in some cases, glass or film copies have been scanned as opposed to
glass originals; note also that some fields have a recent non--atlas plate
or glass copy plate of the original atlas plate substituted where the original
was unmeasurable (e.g.~because of severe degradation due to `microspots' --
see Morgan~1995). All of the scanning programmes listed in 
Section~\ref{intro} have made use of copy material for some areas/passbands.
Although no rigorous quantitative assessment has been done, it is generally
thought that the use of copy material is not significantly detrimental to
the accuracy of the scanned data.
The system of field numbers used by the UK and ESO Schmidt
telescopes is described in Tritton~(1983) and consists of uniformly--spaced
field centres at a $5^{\circ}$ pitch with $\sim0.5^{\circ}$ overlap at all
boundaries (for the UK Schmidt plates); this survey system was reflected into
the northern hemisphere and adopted for the second epoch Palomar Sky Survey
(POSS--II). The first epoch Palomar Sky Survey (POSS--I) used $6^{\circ}$ field
spacing on a different grid (see Minkowski \& Abell~1963) resulting in little
overlap between each field.
\begin{table*}
\begin{center}
\begin{tabular}{lccll}
\multicolumn{1}{l}{Survey} & \multicolumn{1}{c}{Dec} & \multicolumn{1}{c}{Plate} & \multicolumn{1}{c}{Dates of} & \multicolumn{1}{l}{Reference}\\
  & \multicolumn{1}{c}{centres} & \multicolumn{1}{c}{limit} & \multicolumn{1}{c}{observation} & \\
\multicolumn{5}{c}{ }\\
\multicolumn{5}{l}{Southern hemisphere survey:}\\
\multicolumn{5}{c}{ }\\
SERC--J/EJ$^1$ & $\delta\leq0^{\circ}$ & B$_{\rm J}\sim23$ & 1974 to 1994 & Cannon~(1984)\\
SERC--ER/AAO--R$^2$ & $\delta\leq0^{\circ}$ & R~$\sim22$ & 1984 to 2000 & Cannon~(1984); Morgan et al.~(1992)\\
SERC--I  & $\delta\leq0^{\circ}$ & I~$\sim19$ & 1978 $\longrightarrow$ &
Hartley \& Dawe~(1981) \\
ESO--R$^3$ & $\delta\leq-20^{\circ}$ & R~$\sim22$ & 1978 to 1990 &
West~(1984) \\
POSS--I E$^3$ & $-18^{\circ}\leq\delta\leq0^{\circ}$ & R~$\sim20$ & 1949 to 1958 &
Minkowski \& Abell~(1963) \\
\multicolumn{5}{c}{ }\\
\multicolumn{5}{l}{Putative northern hemisphere survey:}\\
\multicolumn{5}{c}{ }\\
POSS--II B$^3$ & $\delta\geq0^{\circ}$ & B$_{\rm J}\sim22.5$ & 1987 to 1999 & Reid et al.~(1991)\\
POSS--II R$^3$ & $\delta\geq0^{\circ}$ & R~$\sim20.8$ & 1987 to 1999 & Reid et al.~(1991)\\
POSS--II I$^4$ & $\delta\geq0^{\circ}$ & I~$\sim19.5$ & 1989 $\longrightarrow$ & Reid et al.~(1991)\\
POSS--I E$^3$ & $\delta\geq0^{\circ}$ & R~$\sim20$ & 1949 to 1958 &
Minkowski \& Abell~(1963) \\
\multicolumn{5}{c}{ }\\
\multicolumn{5}{l}{Notes:}\\
\multicolumn{5}{l}{$^1$Original glass survey plates scanned with the exception
of the following 6 fields: 31, 102, 167, 330, 555,}\\
\multicolumn{5}{l}{ 575 (replacement glass originals).}\\
\multicolumn{5}{l}{$^2$Original glass survey plates scanned with the 
exception of
the following 32 field numbers: 52, 54, 111, 114, 119, }\\
\multicolumn{5}{l}{ 244, 296, 298, 355, 412, 413, 472,  
473, 476, 483, 540, 541, 549, 550, 611, 619, 632, 679, 686, 691, 758, 760, }\\
\multicolumn{5}{l}{ 765, 828, 831, 837, 838 (film originals).}\\
\multicolumn{5}{l}{$^3$Glass atlas copies of survey glass originals.}\\
\multicolumn{5}{l}{$^4$Film atlas copies of survey glass originals.}\\
\end{tabular}
\caption[ ]{Photographic atlases comprising the SSS (after Morgan~1995);
note that, at the time of writing, the I--band surveys are 
together incomplete at a total of $\sim$60 fields.}
\label{atlases}
\end{center}
\end{table*}
A description of the scanning procedure used to digitise the photographic
material is given in Paper~{\sc II} and references therein. 

\subsection{Database organisation and access}
\label{access}

\begin{figure*}
\setlength{\unitlength}{1mm}
\begin{picture}(170,200)
\includegraphics{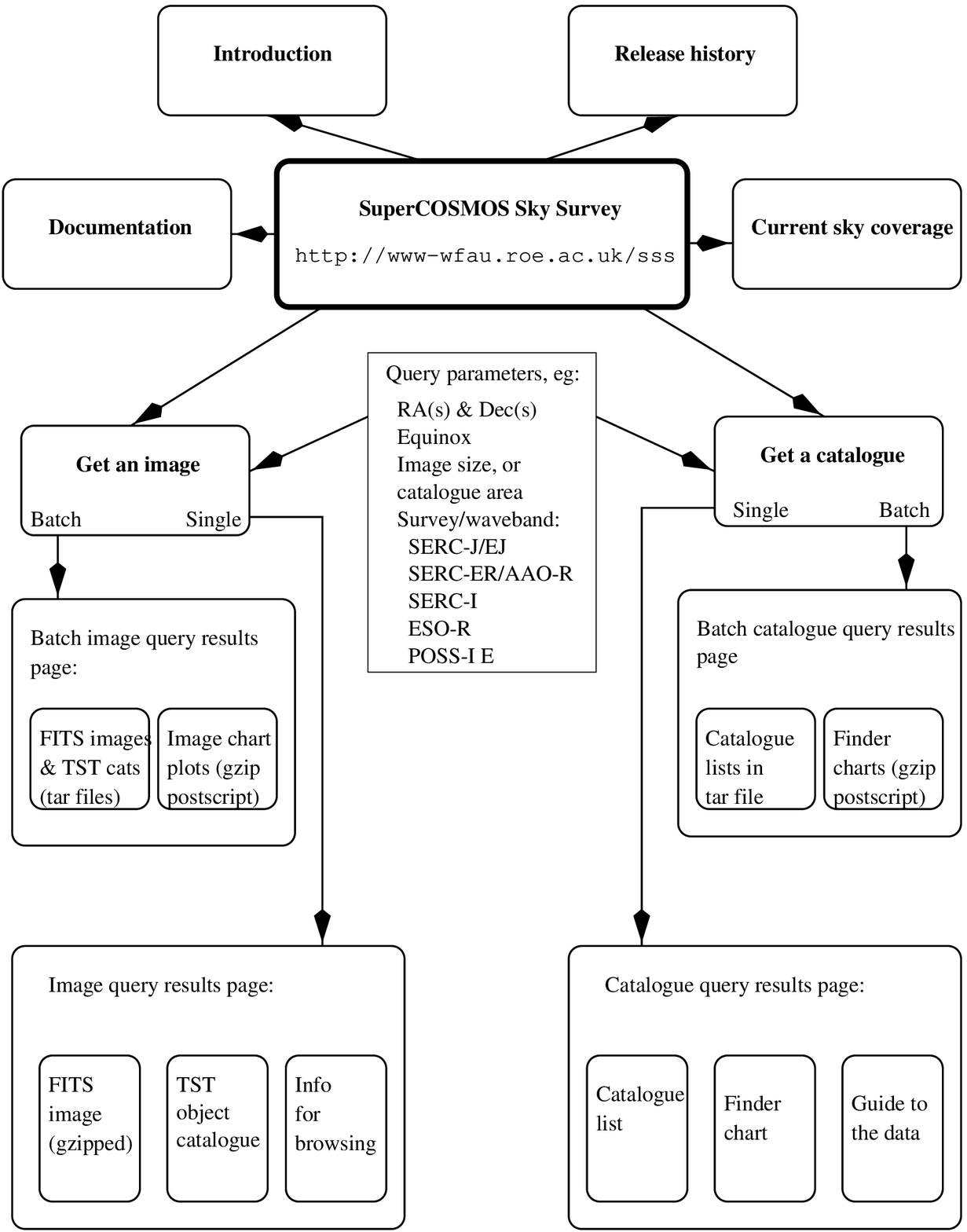}
\end{picture}
\caption[]{Schematic diagram illustrating the user interface to the SSS
database.}
\label{flowchart}
\end{figure*}

SSS data are available in the form of pixel images and/or object catalogues.
Figure~\ref{flowchart} gives a schematic illustration of the user interface,
summarising the main features.
The database organisation and modes of access are quite general, and allow
specification of the colour of interest for image queries, and equivalently
the `master' colour (or colour of most interest) for object catalogue queries.
Object catalogue information from other colours for the same objects/areas
is given for both image and object catalogue queries. The concept of a 
user--specified master (or reference) colour allows an object catalogue
query aimed at studying blue objects, for example, to specify the bluest plates
in the collection as the data of most interest. In this way, objects not
appearing on the redder plates will not be rejected if they are detected on
the blue plates only. Presently, complete lists of objects appearing in all
colours are not generated, although it will of course be possible to do this
once the photography and scanning are complete. For a `cookbook'--style
introduction to access and browsing, the reader is referred to 
Hambly \& Read~(2000).

\subsubsection{Image data}
\label{pixels}

Each digitised Schmidt plate produces $\sim2$~Gbyte of 10~$\mu$m 
($\sim0.7$~arcsec) pixel data (Paper~{\sc II}). A full--sky survey in three
colours, one colour at two epochs results in $>10$~Tbyte of raw pixel data.
In order to be able to store the pixel data {\em and} the associated 
multi--parameter object catalogues online for fast access, we have
employed the H--compress algorithm developed at STScI (White et al.~1992) to
compress the pixel data by a factor $\sim20$. Note that SSS object catalogue 
data are derived from the {\em raw} pixel data before these are archived
to tape for offline storage; however White et al.~(1992) demonstrate the
effect of image compression on astrometric and photometric performance. 
Furthermore, in Figure~\ref{nmcomp} (see Section~\ref{mmcomp}
later) we show how image detection 
performs relative to uncompressed data as a function of
compression factor to show quantitatively the effects of 
using H--compress. Clearly, there is no loss in `depth' as measured
by this test even for the relatively high compression factor of $\times20$.

Pixel data extracted from the SSS database are currently provided in density
units (Paper~{\sc II}, Equation~1) arbitrarily scaled to maximise use of
the range of 16--bit integers per pixel. No photometric calibration is provided
with the images since the dynamic range of the data are image--morphology
dependent, and generally limited to around a factor $\sim2.1$ in density
(e.g.~Hambly et al.~1998) which corresponds to a factor of $\sim10$ in
incident intensity; however approximate relative intensities for pixel values
are given by $I\propto 10^{0.4D}$ where the exponent~0.4 takes into account the
specularity of the SuperCOSMOS imaging optics.

Each image extracted from the database is output in standard 
Flexible Image Transport System format
(see \verb+http://heasarc.gsfc.nasa.gov/docs/heasarc/fits.html+) 
with comprehensive FITS
header information giving details of the source photographic material,
measurement and astrometric calibration. Additionally, FITS World Co-ordinate
System keywords describing the local astrometry are included (see
Paper~{\sc III}).

Each image extracted from the database also comes with an associated object
catalogue so that the image data can be examined in conjunction with the
detected sources in that area (e.g.~by using GAIA/SkyCAT, Draper~1999).
Two object lists are attached as binary table extensions to the image. The
first table consists of the `most useful' subset of all the image
parameters available from the field/colour in question along with proper motion
and colour information from other colours of the same field; the second table
gives the remainder of the full 32 parameters set for each image from the
plate of the specified colour. The equinox of the co-ordinates in object
catalogues associated with images is always J2000.0 while the epoch is always
that of the plate from which the positions are derived (as given in the FITS
image header).

Presently, images up to 30~arcmin in size can be extracted online. Given an
input position, the access software determines from which field the best
image can be obtained (taking into account field overlaps and proximity to
respective field centres).

\subsubsection{Object catalogues}

Image detection and parameterisation are described in Paper~{\sc II}. Each
individual plate's pixel dataset has a raw object catalogue associated with
it; each detected image has 32~parameters describing such quantities as
position, brightness, morphology and image class.
Photometric calibration is described in Paper~{\sc II} while astrometric
calibration is described in Paper~{\sc III}. Object catalogue formats
currently supported in the SSS are ASCII (i.e.\ plain text, the most
human--readable format), the so--called tab--separated table 
(TST, Davenhall~2000 and references therein) and FITS binary table extensions.
Paper~{\sc II} gives details of the internal storage units for object
catalogue data (Table~1, Section~2.1.4). At access time, data are translated
into more conventional units depending on which type of output file is
being written. Note that for the TST and FITS formats, browsers
and manipulation software may translate units internally for display
purposes (for example, SSS FITS binary tables use double--precision floating
point numbers in units of radians for celestial co-ordinates, and these
will often be translated into sexagesimal for display).

\begin{figure*}
\setlength{\unitlength}{1mm}
\begin{picture}(170,220)
\includegraphics{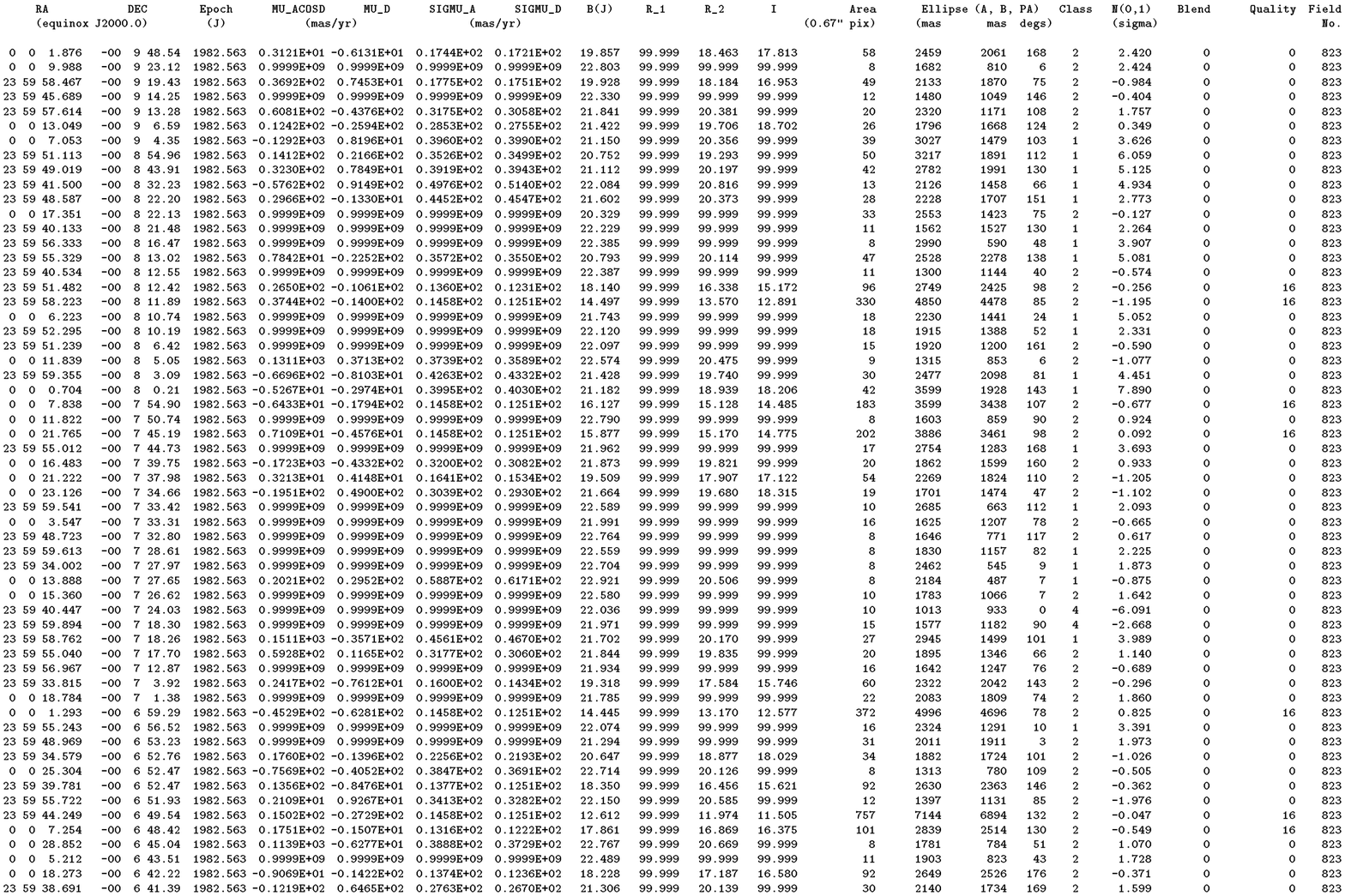}
\end{picture}
\caption[]{Example object catalogue table from an SSS query.}
\label{ltable}
\end{figure*}

ASCII format object catalogues (e.g.\ Figure~\ref{ltable})
consist of object lists with sexagesimal
celestial co-ordinates at the equinox/epoch specified by the user; proper
motion measures and error estimates in units of mas~yr$^{-1}$ (defaulting
to $9.99\times10^8$ if no measurement is available); 
BR$_1$R$_2$I magnitudes in the
natural photographic systems (described, for example, in Bessell~1986 and
Evans~1989 and references therein), defaulting to 99.999 when no measurement
is available in any particular colour; morphological data from the master plate,
including semi--major and semi--minor ellipse--fit axes in mas, celestial
position angle in degrees, classification flag (1=non--stellar, 2=stellar,
3=unclassifiable and 4=noise) and profile classification statistic in units
of sigma from an ideal stellar image (see Paper~{\sc II}, Section~2.2.3);
image blending and quality information; and finally a field number of origin.
TST files give the same information, but the units for the celestial
co-ordinates are decimal degrees, as required by the TST standard. Furthermore,
the ellipse--fit axes are given in units of 10~$\mu$m pixels, where
10~$\mu{\rm m}\equiv0.67$~arcsec. FITS tables again give the same information,
but celestial co-ordinates are stored in units of radians.

\subsection{Object catalogue generation options}
\label{options}

The online server giving access to the SSS data provides a number of options
for object catalogue generation. These fall into the two broad categories of
basic options, and those of a more advanced or expert nature.

Basic user options consist of: search centre and equinox; primary (or `master')
survey waveband, size of extracted area (rectangular or circular);
primary magnitude range; output equinox; output epoch (celestial co-ordinates
can be proper motion corrected to a given epoch, or alternatively left at the
epoch of observation); and finally output format (ASCII, TST or FITS).

More advanced options (which default to sensible values for general queries)
are: i) an option to purge or keep duplicate image in overlap regions; 
ii) subset of
parameters or all~32 per object; iii) a blending option allowing one of four
choices in the presence of blended images; and iv) an image quality threshold.
The image analyser (described in Paper~{\sc II}) flags images according to
detection of multiple components and various quality issues. The blend flag
can be used to discard `parent' images of multiple components (i.e.~the
single image comprising the components) or alternatively the component
(`child') images may be discarded. The other two choices available are to
keep {\em all} images (i.e.~parents and children) or to keep only unblended
images. If completeness is not an issue, then a clean catalogue of isolated,
well defined images can be obtained using this last option. The quality bit
flags are described in Paper~{\sc II}, Table~2 and a given quality threshold
can be specified to discard images, since in general the more significant
quality bits indicate more severe quality issues. For example, once again if
completeness near bright stars is not an issue, a quality threshold of~1023
will reject all objects near a bright star as described in 
Paper~{\sc II}, Section~2.2.2. Since bit~10 in the quality flag is used to
signify proximity to a bright star, and $2^{10}=1024$, then any object
having quality $<1024$ is not affected by this particular quality issue --
i.e.\ is not near a bright star.
Conversely, a quality threshold of~2047 would
{\em include} all those objects (and a good many spurious
images due to diffraction spikes and low--level scattered light halos as well).
The lower the quality threshold, the more
strict the quality assurance is. Note that both blend and quality options are
applied to every plate in creating the object catalogue (i.e.~if other colours
are available in addition to the primary colour, then the quality and blend
tests are applied to their parameters also). All other selection options are
applied to just the primary. Finally, the advanced options also include a
pairing proximity choice of either~3 or 6~arcsec (the latter allows 
investigation of possible high proper motion objects -- Paper~{\sc III},
Section~2.2.1) and a colour correction option allowing the switching on/off
of correction of systematic errors as a function of position/magnitude
(Paper~{\sc II}, Section~2.3.4).

\subsection{`Seamless' catalogues and missing data}

The modern Schmidt atlas surveys have varying degrees of overlap between
adjacent fields. Also, the SSS is an on--going project requiring several 
years to complete the photography and scanning and, if data are to be made
generally available as soon as possible, some data will be unavailable in
specific areas. Moreover, paired information (i.e.~colours and proper
motions) are not always available if, for example, an image appears in
one passband only. For all these reasons, a decision has to be made as to
what to do in the case that more than one set of records, or no records
at all, are available for given images or areas.

For the case of duplicate images in overlap regions the SSS database access
software uses proximity to respective plate centres combined with quality
information to decide which image to keep in the case that duplicates
exist. Briefly, out of the list of duplicates available (which can be up to
four in the extreme field corners), any having quality indicating proximity
to a plate label or wedge (Paper~{\sc II}, Section~2.2.2) are discarded and
then the object nearest the respective field centre is chosen as the `best'
image. Note that `parent' images are unpaired in the database and will not
be purged in overlap regions if selected. 

For situations where paired information is not available (because the
required data have not yet been included in the database, or because no
image pairs with other colours exist) then null values are output for
proper motions and colours. These are $9.999\times10^8$ for proper motion
components and errors and $99.999$ for magnitudes. Of course, in the case
that no primary image records are available in a requested area, then
nothing will appear in the extracted object catalogue dataset.

\section{Results and discussion}
\label{results}

\subsection{Comparison with data from other scanning programmes}
\label{mmcomp}

\subsubsection{Pixels}
\label{hcomp}

For a quantitative comparison of the STScI and SuperCOSMOS pixel data products,
a $30\times30$~arcmin region in ESO/SERC field~411 (i.e. at the South Galactic
Pole) was chosen for test purposes. First (J) and second (R) generation DSS
data were retreived from the STScI data server. These data are compressed
by a factor $\sim10$ to facilitate on--line storage at STScI; the SuperCOSMOS
data were compressed using H--compress (White et al.~1992) by
factors of~10 and~20 to compare directly with DSS data and also to show the
effects of using higher compression factors. In order to assess quantitatively
and objectively the information content of the images, the data were run 
through PISA (Draper \& Eaton~1999) which is
an isophotal pixel analysis package similar to
the COSMOS image analyser (Beard, MacGillivray \& Thanisch~1990) that
both ROE and STScI use to generate object catalogues (note of course that 
both groups provide object catalogues derived from uncompressed data;
running decompressed images through object detection software is done here
for illustrative purposes only). The data were thresholded (in 
density space)
at $2.3\sigma$ above the mean sky level (Paper~{\sc II}, Section 2.1.4) and a
minimum of 6~pixels was required to define an image in all cases. Note that
this minimum pixel area cut was kept the same for the tests despite the fact
that the STScI data have larger pixel sizes ($25\mu$m and $15\mu$m for first
and second generation data respectively), since reducing the cut for 
the coarser sampled data to an area equivalent to that of the higher 
resolution SuperCOSMOS data results in large numbers of spurious noise 
images being detected --- it is
fair to say that the fact that the SuperCOSMOS data has higher spatial
resolution allows fainter detection limits, and this should be reflected in
any test.

\begin{figure*}
\setlength{\unitlength}{1mm}
\begin{picture}(170,120)
\includegraphics{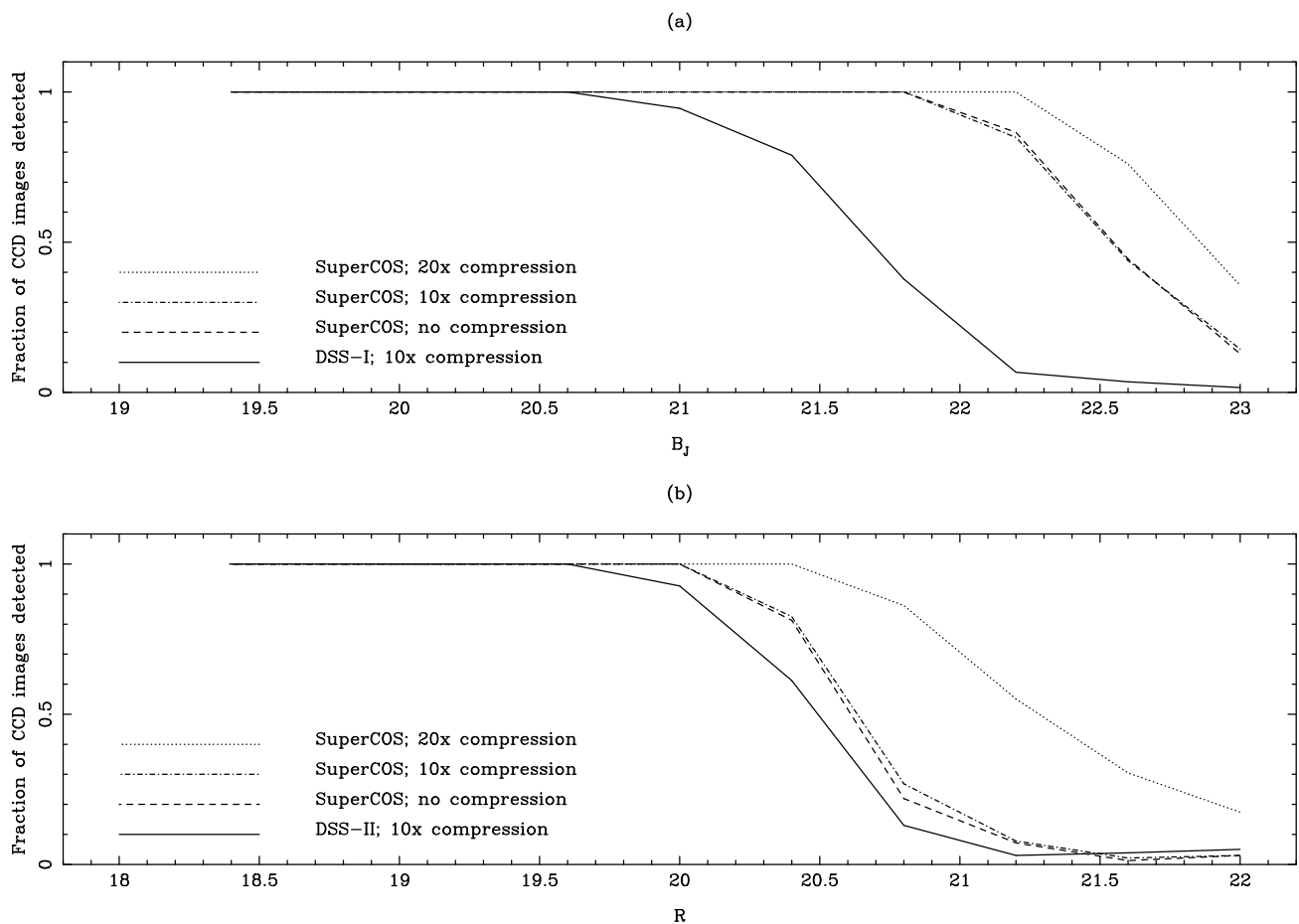}
\end{picture}
\caption[]{Number--magnitude histograms for 
(a) DSS--I, and (b) DSS--II with corresponding
SuperCOSMOS data compared against deep CCD data in the SGP.}
\label{nmcomp}
\end{figure*}

Figure~\ref{nmcomp} 
shows the results of the image detection tests when compared
against a CCD image of the same field which reaches much fainter limits than
the photographs. From the plots, it can be seen immediately that the 
SuperCOSMOS pixel data are superior (in this test, they reach $\sim1$~mag
deeper) than the DSS--I data, while the DSS--II and SuperCOSMOS data are broadly
similar. Interestingly (and counter--intuitively), the image detection is
deeper for higher compression factors. This has been noted before 
(e.g.~White et al.~1992) and can be understood in terms of the compression
algorithm smoothing the background which results in a lower detection
threshold when computing $n\sigma$. 

\subsubsection{Astrometry}

In Paper~{\sc III}, we demonstrate the external accuracy of the SSS data
for $|b|\geq30^{\circ}$
using the test described by Deutsch~(1999) based on standard positions from
the International Celestial Reference Frame (ICRF) sources listed in
Ma et al.~(1998). Table~\ref{astcomp} summarises the results from 
Deutsch~(1999) along with similar results from Paper~{\sc III}. In the Table,
the columns $<\Delta\alpha>$, $<\Delta\delta>$ give the average offsets of the
survey data from the ICRF sources; $N$ is the number of sources used while
$\Delta\alpha$, $\Delta\delta$ and $\Delta R$ are estimates of the scatter in
the distributions in RA, Dec and radially (see Deutsch~1999 for more details).
\begin{table*}
\begin{center}
\begin{tabular}{cccccccc}
Database & $<\Delta\alpha>$ & $<\Delta\delta>$ & $N$ &  $\Delta\alpha$ & $\Delta\delta$ & $\Delta R$ & Note \\
         & \multicolumn{2}{c}{(arcseconds)} &        &  \multicolumn{3}{c}{(arcseconds)} & \\
\multicolumn{8}{c}{ }\\
SSS~(J)          & $-0.05$ & $-0.04$ & 110 & 0.12 & 0.12 & 0.17 & 1 \\
SSS~(R)          & $-0.09$ & $+0.07$ & 103 & 0.14 & 0.23 & 0.27 & 1 \\
USNO--A2.0       & $+0.04$ & $+0.04$ & 283 & 0.17 & 0.18 & 0.26 & 2 \\
DSS--II$(+1,-1)$ & $-0.26$ & $+0.07$ & 235 & 0.63 & 0.52 & 0.85 & 3 \\
DSS--I           & $-0.08$ & $+0.03$ & 274 & 0.53 & 0.48 & 0.75 &   \\
A2.0$\rightarrow$DSS--I &
                   $+0.02$ & $+0.03$ & 104 & 0.18 & 0.19 & 0.32 & 4 \\
A2.0$\rightarrow$DSS--II &
                   $+0.04$ & $+0.13$ & 108 & 0.21 & 0.26 & 0.35 & 4 \\
APM (UKST)       & $-0.06$ & $+0.06$ & 126 & 0.18 & 0.22 & 0.28 & 5 \\
APM (POSS--I)    & $-0.05$ & $+0.05$ & 231 & 0.27 & 0.31 & 0.41 & 5 \\
\multicolumn{8}{c}{ }\\
\multicolumn{8}{l}{Notes:}\\
\multicolumn{8}{l}{1: SSS data from Paper~{\sc III}}\\
\multicolumn{8}{l}{2: USNO--A2.0 are the {\em average} of positions from J \& R plates}\\
\multicolumn{8}{l}{3: results allowing for a 1 pixel offset in DSS--II image WCS
(see Deutsch~1999)}\\
\multicolumn{8}{l}{4: results from using USNO--A2.0 to rereduce astrometry for
DSS images}\\
\multicolumn{8}{l}{5: data from 
\verb+http://www.ast.cam.ac.uk/+$\sim$\verb+mike/apmcat+}
\end{tabular}
\caption[ ]{Empirical uncertainty estimates in survey astrometry relative to
the ICRF (after Deutsch~1999; see also Paper~{\sc III}).}
\label{astcomp}
\end{center}
\end{table*}

Comparing figures in Table~\ref{astcomp}, the SSS data are as good as, or better
than, all the other survey data. It should perhaps be emphasised that a mixture
of reference catalogues have been used to astrometrically reduce the survey
datasets used in Table~\ref{astcomp} -- e.g.~Tycho--2 for APM and SSS data
(Paper~{\sc III}); Tycho--AC for USNO--A2.0. Nonetheless, at the time of
writing, these data were the best available. 

Direct comparison of proper motion errors between the various 
surveys described in
Section~\ref{intro} was not possible at the time of writing due to the
unavailability of the GSC--II and USNO--B catalogues. SSS proper motions are
determined to be typically accurate to $\sim10$~mas~yr$^{-1}$ in an absolute
sense at m~$\sim18$ (e.g.\ Paper~{\sc III}).
GSC--II proper motions are expected to be accurate to
better than 4~mas~yr$^{-1}$ (McLean et al.\ 1997b); however it remains to be
seen whether this is possible in the southern hemisphere, and whether it is a 
true absolute error (i.e.\ including any zeropoint errors) or merely a
relative error under the most favourable conditions. As yet, there is no 
indication as to proper motion accuracy for USNO--B.

\subsubsection{Photometry}

Comparisons of catalogue photometry are even less clear than those for astrometric
parameters. The absolute accuracy of photographic photometry will always be
limited by position-- and magnitude--dependent errors (Paper~{\sc II}); however
as we have shown in the same paper, colours can be adjusted to remove such
systematic errors from indices such as (B$_{\rm J}-{\rm R}$) or 
$({\rm R}-{\rm I})$. Once again, it is as yet unclear whether such 
procedures will be used for many of the other surveys detailed 
in Section~\ref{intro}. In broad terms,
GSC--II photometry is predicted to be accurate to 0.1--0.2~mag whereas GSC--I 
was only good to 0.4~mag (McLean et al.~1997b); an idea of the accuracy of colours
for DPOSS may be obtained from Kennefick et al.~(1995) -- e.g.~their Figures~1
compared with Paper~{\sc II}, Figure~10. SSS data are accurate in an 
{\em absolute}
sense to $\sigma_{\rm B,R,I}\sim0.3$ (Paper~{\sc II}). However we note that
SSS colours (e.g. B$_{\rm J}-{\rm R}$) are much more precise: 
$\sigma_{\rm B-R}\sim0.07$~mag at B$_{\rm J}\sim17$ rising to 
$\sigma_{\rm B-R}\sim0.16$~mag at B$_{\rm J}\sim20$. 
Colour indices are not susceptible to systematic errors (which dominate the
single--passband precision of 0.3~mag) due to the calibration procedure
described in Paper~{\sc II}, Section 2.3.4.
The USNO catalogues are primarily for astrometric purposes, so a comparison of
their photometric properties is inappropriate.

\subsection{Science examples using SSS data}

Applications of the SSS data are many and varied, ranging from the small--scale
(e.g.\ finding--chart 
generation, secondary astrometric standards for deep images
from other instruments/wavebands) through intermediate angular scales (e.g.\
statistical studies of the clustering properties of galaxies over degree scales)
to the widest areas (e.g.\ statistical studies of the properties of galaxies
averaged over large fractions of the whole sky).

Examples of small--scale applications of SSS data include optical identification
of sources and registration of images from different wavebands (e.g.\ Galama et 
al.\ 1998; Jonker et al.\ 2000; Stairs et al.\ 2001). Degree--scale
examples include identification of stellar samples from photometry (e.g.\
Preibisch \& Zinnecker~1999) and the use of survey plate astrometry to provide a
precise astrometric reference frame for fibre spectroscopy, for example open
cluster work in IC~2391 (Barrado--y--Navascu\'{e}s et al.\ 2001) and
NGC~2547 (Jeffries et al.\ 2000). In the remainder of this Section, we discuss 
some results from wide--angle studies employing SSS data.

\subsubsection{Intrinsic alignments of galaxies at low redshift}

The SGC subset of the SSS data has been used to attempt to
detect the large scale structure of the universe via distortions in the
shapes of background galaxies caused by gravitational lensing by the
intervening mass distribution. It turns out that the variance of
galaxy ellipticities, $\sigma^2_e$ over some angular scale is a
measure of this `cosmic shear' effect over the same
scale assuming that the background galaxies are initially randomly
orientated (e.g.\ Kaiser~2000 and references therein).
If this is not the case and there is some intrinsic
preferential alignment of the background galaxies, then $\sigma^2_e$
will give an estimate of the size of this intrinsic alignment.
Brown et al.~(2001) have shown that,
for a magnitude cut of B$_{\rm J}<20.5$ corresponding
to a median galaxy redshift of z~$\approx0.1$, the galaxy
ellipticities exhibit a non--zero correlation over scales between~1 and~100
arcmin. The variance of mean galaxy ellipticities is $\sim10^{-2}$ 
at scales of $\sim10$ arcmin and falls rapidly to $\sim10^{-4}$ at scales of
$>30$ arcmin. Because of the low median redshift of the galaxies in the sample
and the size of the signal it is concluded that the intrinsic 
alignment of galaxies has been measured, 
rather than the `weak' lensing signal due to dark matter.

\subsubsection{A survey for candidate Halo cool white dwarfs}

Recently, there has been renewed interest in the possibility that a significant
fraction of the Galactic Halo dark matter may be in the form of cool, ancient
Halo white dwarfs (eg.\ Hodgkin et al.~2000; Ibata et al.~2000
and references therein). Wide--field,
multi--epoch Schmidt data are ideal for searching for such objects, since they
should have large intrinsic space motions and therefore large proper motions.
The SGC subset of the SSS data 
have been analysed for high proper motion stars using the B and
two R epochs to provide a clean, three--epoch sample with quantifiable 
completeness (see Hambly~2001 for more details). Follow--up spectroscopy on
a large sample of objects subluminous in reduced proper motion has uncovered
a significant number of new cool white dwarfs; a preliminary analysis
indicates that the contribution to the `standard' dark matter Halo from these
objects is at least $\sim2$\% (Oppenheimer et al.~2001).

\subsubsection{Galaxy two--point angular correlation}

In principle a measurement of the two-point correlation function of
galaxies $\xi(r)$ can provide useful constraints on the cosmological parameters
which govern the evolution of the universe. Also an understanding of
its evolution with redshift and morphological type (roughly colour
selected sub--samples) can reveal the required ingredients for consistent
models of galaxy formation (Peacock \& Smith~2000, Benson et al.~2000).
In the absence of any distance information, one can compute the
two--point angular correlation function $\omega(\theta)$ of the
sky--projected distribution of galaxies, which is defined as the excess
probability above a Poissonian distribution for finding a galaxy in
the angular areas $d\Omega_1$ and $d\Omega_2$ separated by an angular
scale $\theta$; explicitly
$dP_{12}=\bar{n}^2[1+w(\theta_{12})]\ d\Omega_1\ d\Omega_2$. 
Although a less sensitive statistic than $\xi(r)$, $\omega(\theta)$
does not suffer from redshift space distortions.

We have measured $\omega(\theta)$ for a sub--sample of the SSS; the
sub--sample is enclosed within a circular region around the SGP 
covering roughly 3400 square
degrees and contains over three million galaxies to a depth of
B$_{\rm J}=21$. The correlation estmates were computed using the simple
statistical estimators
$\hat{\omega}(\theta)_1 = 1 + <DD>/<RR>$
and
$\hat{\omega}(\theta)_2 = 1 + <DD>/<DR>$
where $<DD>$ represents the expected number of measured galaxy--galaxy pair
counts, $<RR>$ the expected number of random--random pair counts and
$<DR>$ the expected number of galaxy--random pair counts in the angular
range $\theta\rightarrow\theta+d\theta$. 

The resultant measurements of $\omega(\theta)$ for four
magnitude slices of thickness $0.58$ in the range $17.0<{\rm B}_{\rm J}<19.3$
have been computed assuming Poissonian errors. The
amplitude of $\omega(\theta)$ increases with brightness as expected.
Quantitavely, a simple linear least squares fit to the power-law slope
on scales $< 1$ degree gives for bright to faint slices
$\delta=0.57,\:0.61,\:0.56$ and $0.59$, with errors of order $\pm
0.02$; where $\omega(\theta)\propto\theta^{-\delta}$. This absence of
evolution in $\delta$ affirms the angular clustering meaurements from
the APM survey (Maddox et al.~1990).

\begin{figure}
\setlength{\unitlength}{1mm}
\begin{picture}(80,70)
\includegraphics{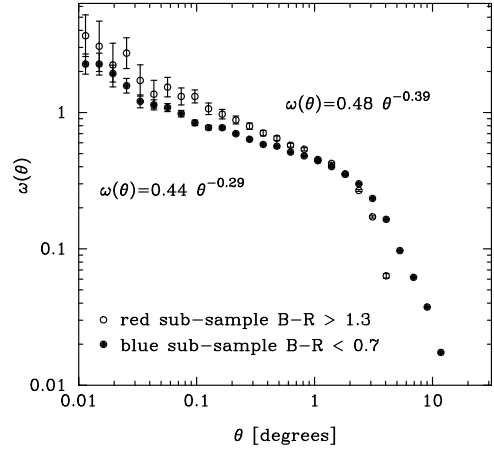}
\end{picture}
\caption[]{Galaxy two--point correlation function computed for red and blue
galaxy samples within the SGC survey (see text).}
\label{wtheta}
\end{figure}

In Figure~\ref{wtheta} we compare the angular clustering for red and blue,
colour selected sub--samples in the magnitude range 
$17.0<{\rm B}_{\rm J}<18.0$. 
We define blue galaxies to be those with ${\rm B}_{\rm J}-{\rm R} < 0.7$ and red
galaxies to be those with ${\rm B}_{\rm J}-{\rm R} > 1.3$; 
the median ${\rm B}_{\rm J}-{\rm R}$ for the survey
is 1.1. We find that on scales $<1^{\circ}$, red galaxies cluster more
strongly than blue and that on scales $>2^{\circ}$, blue galaxies are
more strongly correlated (cf.\ Brown, Webster \& Boyle~2000). 
Our interpretation of these two results is
that red galaxies trace the clusters and since these are high peaks
in the density field cluster more strongly on small scales; whereas
blue galaxies have a higher propensity to be found in the field, so
trace the large filamentary structures. Further details of this work will
appear in Smith et al.~(2001).

\subsubsection{A survey for objects exhibiting extreme variability}

\begin{figure}
\setlength{\unitlength}{1mm}
\begin{picture}(80,60)
\includegraphics{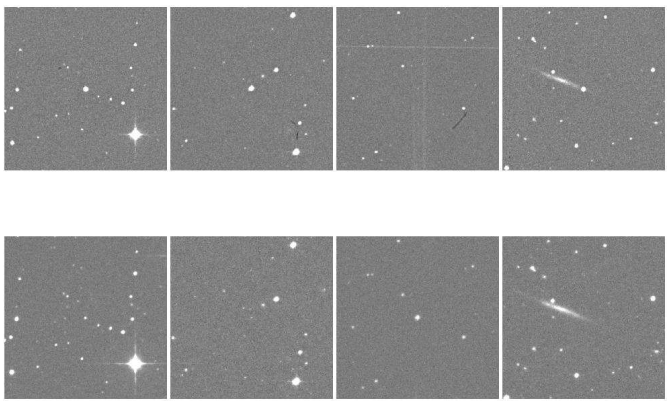}
\end{picture}
\caption[]{Four examples of objects showing extreme variability
between the SSS R--band observation epochs ($\Delta{\rm m}>5$).}
\label{rvars}
\end{figure}

The SSS survey contains R--band data at two epochs, enabling investigation
of object variability. Figure~\ref{rvars} shows four examples
of sources showing extreme variability obtained using strict
selection criteria (e.g.\ in image shape and blending flag) from the SGC subset.
The first three objects are presumably
Galactic stellar variables ($\Delta{\rm m}>5$)
and are possibly Miras or Novae; the final example is apparently an old
supernova (m$_{\rm R}\sim14$; plate epoch 1983.8)
associated with the edge--on Scr spiral ESO~148--20. Checks in the database
archive SIMBAD yield no known catalogued objects at these positions, so these
are apparently new discoveries; many more are almost certainly present in the
SSS data. 

\subsubsection{Number--magnitude counts for galaxies}

\begin{figure}
\setlength{\unitlength}{1mm}
\begin{picture}(80,60)
\includegraphics{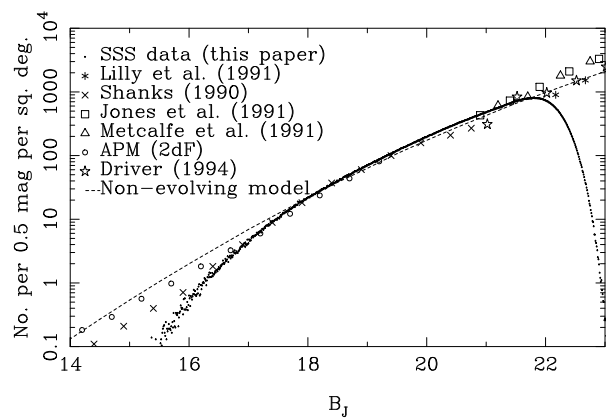}
\end{picture}
\caption[]{Number--magnitude counts for galaxies from the full SSS~J 
data at Galactic latitudes $|b|>45^{\circ}$ along with comparison
data from other surveys.}
\label{gcounts}
\end{figure}

In Figure~\ref{gcounts} we
show a recent compendium of galaxy counts spanning from the 
very local photographic surveys (e.g.\ SSS and APM) to 
deep CCD--based pencil--beam surveys. The SSS counts were made using all
objects classed as galaxies and having $|b|>45^{\circ}$ in the full
southern hemisphere survey. We extinction--corrected all galaxy magnitudes 
using $A_{{\rm B}_{\rm J}}=4.035E_{\rm B-V}$ where $E_{\rm B-V}$ was 
derived from the maps of Schlegel, Finkbeiner \& Davis~(1998).
A seamless catalogue was created with
quality threshold set at 127 and deblended images were excluded
(see Section~\ref{access}). The SSS
data agree very well with existing data. The dashed line shows 
the expected counts based on the most recent 2dF 
galaxy redshift luminosity function, the assumption of a flat 
zero--lambda cosmology and neglecting evolution. The model is 
normalised to the data at B$_{\rm J}=19$. In general the model and 
data agree over the magnitude range $17.5 < {\rm B}_{\rm J} < 21.5$. 
In detail the
photographic data show some discrepancy from B$_{\rm J}\sim20$ indicating 
the onset of selection bias. This is most likely due to increasing numbers
of galaxies being unresolved on the plates (and therefore being classified
as stars) and also the effect 
of surface--brightness dimming where some fraction of the distant
population will be dimmed below the detection isophote. At the brightest
magnitudes, the image detection algorithm `breaks up' galaxy images leading
to the underprediction with respect to the model and other surveys.

There are $1.65\times10^7$ galaxies in the catalogue used for this
 plot -- the SSS data in Figure~\ref{gcounts} are binned per 0.01~mag,
 normalised to 0.5~mag bin width. The smoothness of the curve demonstrates
 the extremely low sampling noise. 

\subsection*{ }

The web--based interface described in Section~\ref{access} allows object
catalogue generation over areas up to $\sim100$ square degrees. In this
Section, we have described some science that is currently being done using
local access to large area object catalogues comprising essentially 
{\em all} of the SSS catalogued data. It is envisaged that provision for
web--based access to the whole database for such statistical analyses will
be implemented in the medium term (see Section~\ref{concs}).

\section{Limitations of the data}

As is the case with any real imaging system, the procedure of automatic analysis
of machine scans of photographs gives rise to various image and object
catalogue artefacts that are spurious and yet may be mistaken for real
phenomena in certain situations. Moreover, the pixel analysis and calibration
procedures (Papers~{\sc II} and~{\sc III}) have their limitations; there are
demonstrable systematic errors in some astrometric and photometric parameters
in {\em all} data derived from photographic plates (e.g.\ the surveys described
in Section~\ref{intro}). In this Section, we summarise some of these limitations
and systematic errors in the SSS
(some of which are described in more detail elsewhere in
this series of papers) and in some cases describe techniques to minimise their
impact when using the data.

\subsection{Known systematic errors in photometry}

As illustrated in Paper~{\sc II}, the absolute accuracy of the photometric
calibration of any given passband are typically $\sigma\sim0.3$~mag for
m~$>15$. At the plate limit, this error is emulsion noise dominated and is
essentially random (i.e.\ normally distributed). For magnitudes brighter than
about 3~mag above the respective plate limits, the precision is limited by
position-- and magnitude--dependent systematic errors. For magnitudes brighter
than m~$\sim14$, systematic errors can become larger than $\sim0.5$ mag depending
on passband, field and position within the field. Note however that systematic
errors in colours are eliminated using the technique described in 
Paper~{\sc II}, Section~2.3.4.

Note that the photometric calibration for stars and galaxies is different, and 
that the classification flag is used to decide which calibration and magnitude
measure (isophotal or profile) are used when accessing the data. At the faint
end where classification reliability is most likely to be poor the 
calibrations converge so systematic errors in photometry from incorrect
classification will be negligible. However, for bright images, incorrect
classification (e.g.\ because of deblending errors) can produce incorrect
magnitude calibration.

\subsection{Known systematic errors in astrometry}

In Paper~{\sc III} we show that the positional absolute astrometric accuracy is
limited by field position dependent systematic errors for most images, resulting
in errors with respect to the ICRF of typically $\sim0.2$~arcsec with some
dependence on passband and field position. A very small number of very low
Galactic latitude~B$_{\rm J}$ and~R plates have absolute positional errors up 
to $\sim3\times$ worse than this due to extreme crowding of the Tycho--2 astrometric
standards. Such fields may be identified by examining the FITS headers for
extracted images. Two keywords (\verb+ASTSIGX+ and \verb+ASTSIGY+) are written
with the WCS keywards which give the average residual of the Tycho--2 standards
used in the global field astrometric solution (see Paper~{\sc III}); these
figures may be taken as an indication of the likely level of any systematic
errors within the field in such cases. Note that on scales of $\sim0.5^{\circ}$
or less, and provided a field boundary is not being crossed, the systematic
errors are constant so relative accuracy is typically better than 0.1~arcsec.

Proper motion errors are dominated by magnitude--dependent systematic errors 
for images brighter than m~$\sim18$ (Paper~{\sc III}). For a typical 15~yr
baseline, the absolute accuracy of the proper motions is 
$\sigma\sim10$~mas~yr$^{-1}$ in either co-ordinate. 

\subsection{Detection completeness and false detections at faint magnitudes}

An indication of the detection completeness of SSS data is demonstrated in
Paper~{\sc II}. With respect to deeper external data, SSS image catalogues are
$>90$\% complete to within $\sim1.5$~mag of the respective plate limits. For the
SERC--J/EJ survey, this gives $>90$\% completeness to B$_{\rm J}\sim20.5$, and
for the SERC--ER/AAO--R the same level of completeness to R~$\sim19.5$.

The question of noise detections at the plate limit is as yet
unmeasured and is difficult to quantify theoretically. Although images are
rigorously required to have 8 connected pixels with intensities $\geq2.3\sigma$
above local sky (Paper~{\sc II}), the pixel--to--pixel noise is correlated to
varying degrees in the image data. All that can be said here is that not every
faint image will be real, and if clean and reliable datasets are required then
it is advisable to use datasets paired between two or more colours when using
SSS data. In any case, great care should be taken when using data within 
$\sim1$~mag of the respective plate limits.

\subsection{Limitations of deblending}
\label{gbreak}

\begin{figure*}
\setlength{\unitlength}{1mm}
\begin{picture}(160,80)
\includegraphics{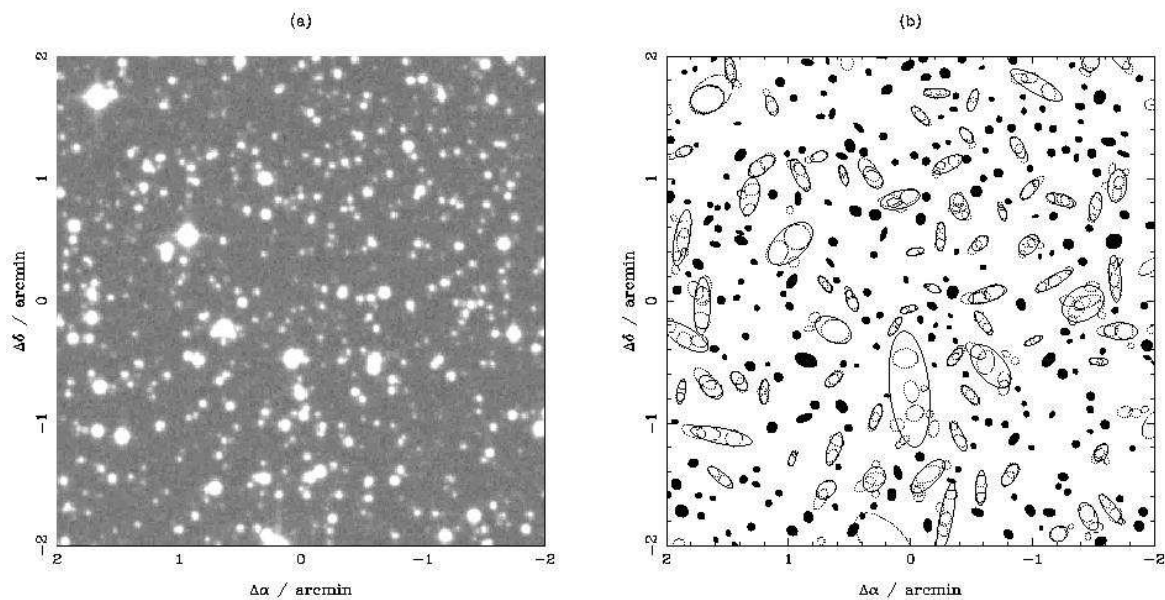}
\end{picture}
\caption[]{Illustration of the deblending algorithm at Galactic co-ordinates
$(l,b)=(0.0^{\circ},-15.0^{\circ})$: (a) pixel data; (b) resulting image 
detections. Filled shapes are isolated images or images not detected as
blends by the algorithm; solid line ellipses
are parent images of blends while the
deblended child images are shown as dotted ellipses. When faint images are 
blended with bright images, the image parameters can be adversely affected
(see Beard et al.~1990).}
\label{deblend}
\end{figure*}

Figure~\ref{deblend} shows an image and corresponding object catalogue ellipse
plot from a $4\times4$ arcmin field at $(l,b)=(0.0^{\circ},-15.0^{\circ})$,
demonstrating some limitations of the deblending algorithm (which are
discussed in some detail in Beard et al.~1990).
Briefly, deblending can only take place when rethresholding detects the
fragmentation of an image; when a faint image is deblended from a bright image,
object parameterisation (particularly the second moments) for the faint image 
can be badly affected. If completeness is not an important issue, then it is
of course possible to use the deblending flag to choose only isolated images
(e.g.\ Section~\ref{options}). Figure~5 of Beard et al.~(1990) demonstrates the
relative completeness as a function of image density when deblending is/is not
employed. At object densities of $\sim20$~arcmin$^{-2}$ (corresponding to
$|b|\sim10^{\circ}$ for example) completeness is $\sim55$\% without deblending
whereas it is $\sim75$\% when deblended images are used.

\begin{figure*}
\setlength{\unitlength}{1mm}
\begin{picture}(160,70)
\includegraphics{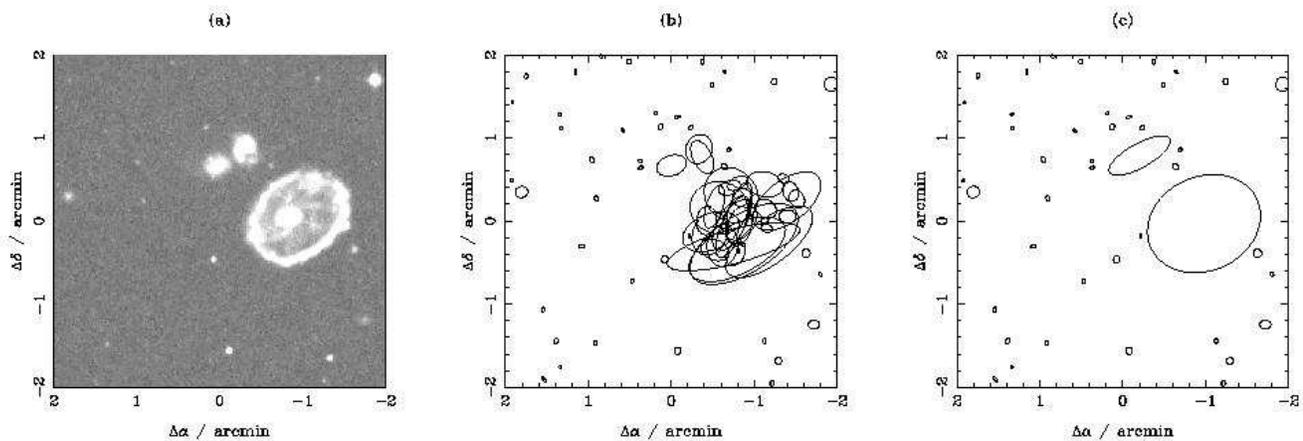}
\end{picture}
\caption[]{Example of image `break--up' from deblending of the relatively bright
Cartwheel galaxy: (a) J--band pixel data; (b) image detections employing
deblended images; and (c) image detections with deblended images ignored.}
\label{gblend}
\end{figure*}

For bright galaxy work, it is important to note that the image deblending
algorithm tends to `break up' such extended, structured images. In 
Figure~\ref{gblend} we show an example for the Cartwheel galaxy
($\alpha\sim0^{\rm h}37^{\rm m};\delta\sim-33^{\circ}43$). If bright resolved
galaxies are of interest, then it may be advisable to ignore deblended images,
and include only those images for which blend~$\leq0$; however it should be 
noted that at present, parent images in the SSS database are unpaired, so
colour information and `seaming' in plate overlap regions are not available.

\subsection{Image classification}

Again, Paper~{\sc II} quantifies the reliability of image classification with
respect to external datasets -- a typical figure is $>90$\% reliability of
correct classification to B$_{\rm J}\sim20.5$. However, it is important to
note that two classification
parameters are available for each image in each passband in SSS data: the
discrete classification code or flag, and the normalised profile 
classification statistic $\eta$. The discrete code is defined as 4 (noise)
for $\eta<-3.0$, stellar for $-3.0\leq\eta\leq+2.0$ and non--stellar for
$\eta>+2.0$. Hence, the classification flag is conservative in the sense of
defining a clean but incomplete stellar sample. Some applications may require
a less strict cut on image class (or a more strict cut on non--stellar class),
in which case it may be advisable to use $\eta$ as opposed to the flag to
define, for example, a stellar selection having $-4.0\leq\eta\leq+4.0$. This
is equivalent to a k--sigma cut on the classification statistic, with
corresponding confidence levels on stellar completeness assuming $\eta$ is
normally distributed for stars.

\subsection{Image defects and spurious objects}

Finally, we address the issue of image defects and spurious objects. As stated
before, the combination of the photographic emulsion, digitisation and
subsequent SuperCOSMOS pixel analysis is by no means a perfect imaging system.
The following defects may be present in SSS data (more description of some of
these may be found in Tritton~1983):

\subsubsection{Emulsion microspots (very rare)}

\begin{figure*}
\setlength{\unitlength}{1mm}
\begin{picture}(160,70)
\includegraphics{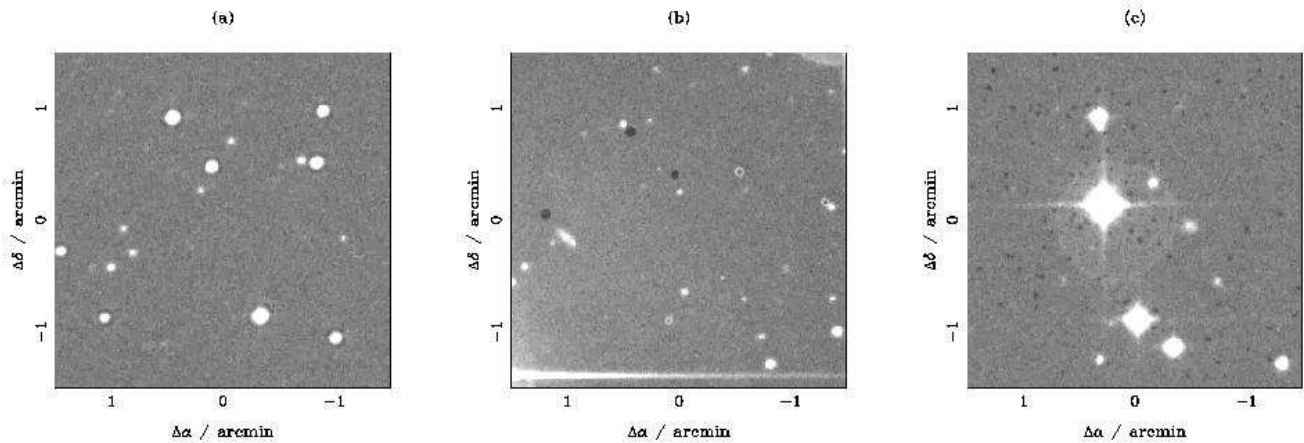}
\end{picture}
\caption[]{Some examples of emulsion microspots (artefacts darker than the
sky in these images) in SSS imaging data (SERC--EJ):
(a) `gold rings' around brighter images; (b) `yellow spots' near a bright
image; and (c) an unusually severe case of `yellow dots'.}
\label{spots}
\end{figure*}

Figure~\ref{spots} shows some examples of image data containing microspots, or
emulsion flaws which appear over time on stored photographic material. These
include classic `gold spot', `gold rings' and `yellow spots'. Many of these
flaws tend to be associated with higher density regions in the developed
emulsion (e.g.\ around bright stars) and nearer the edges of the photographs.
Where spotting was severe over large areas of the original atlas plates we
have substituted a copy plate (made soon after the original was taken and
developed) -- see Table~\ref{atlases}. Hence, the online digital data may
differ from previously issued film atlases for example. Note that in the
interests of a uniform survey we have limited substitution of originals by
copies, and note also that where any spotting is present at the edges of
fields, the adjacent plate overlap object data are generally available.
Moreover, as shown in Figure~\ref{spots}, because microspots for the most part
tend to produce
higher transmission values than the sky level (i.e.\ apparent
emulsion `holes') they do not give rise to spurious object detections. 
However, one effect that has been noticed in the data is that bright images
affected by `gold rings' tend to be flagged as noise images by the image
classifier, so in the very rare instances that microspots are present, image
parameters may be affected.

\subsubsection{Satellite trails (fairly common)}

As an inevitable consequence of long exposures over wide angles in the modern
age, many satellite trails are present on Schmidt atlas photographs. Moreover,
trails resulting from meteors and other atmospheric phenomena can be present.
Under pixel analysis, trails tend to fragment into a series of faint, elliptical
images classed as galaxies. The easiest way to eliminate such spurious detections
from an SSS catalogue is to use a paired dataset -- for example, Brown et 
al.~(2001) used a B$_{\rm J}$/R 
paired galaxy catalogue (i.e.\ a galaxy catalogue made by
using the B$_{\rm J}$ plates as the master but in subsequent analysis 
only images with
a corresponding R magnitude were used) despite the fact that the statistics were
computed from the B$_{\rm J}$ data alone. 
For stellar work, an ellipticity cut can go
a long way to eliminating spurious objects resulting from trails (as well as 
those from other sources -- e.g.\ see the next Section).

\subsubsection{Spurious images around bright stars (very common)}

\begin{figure*}
\setlength{\unitlength}{1mm}
\begin{picture}(160,210)
\includegraphics{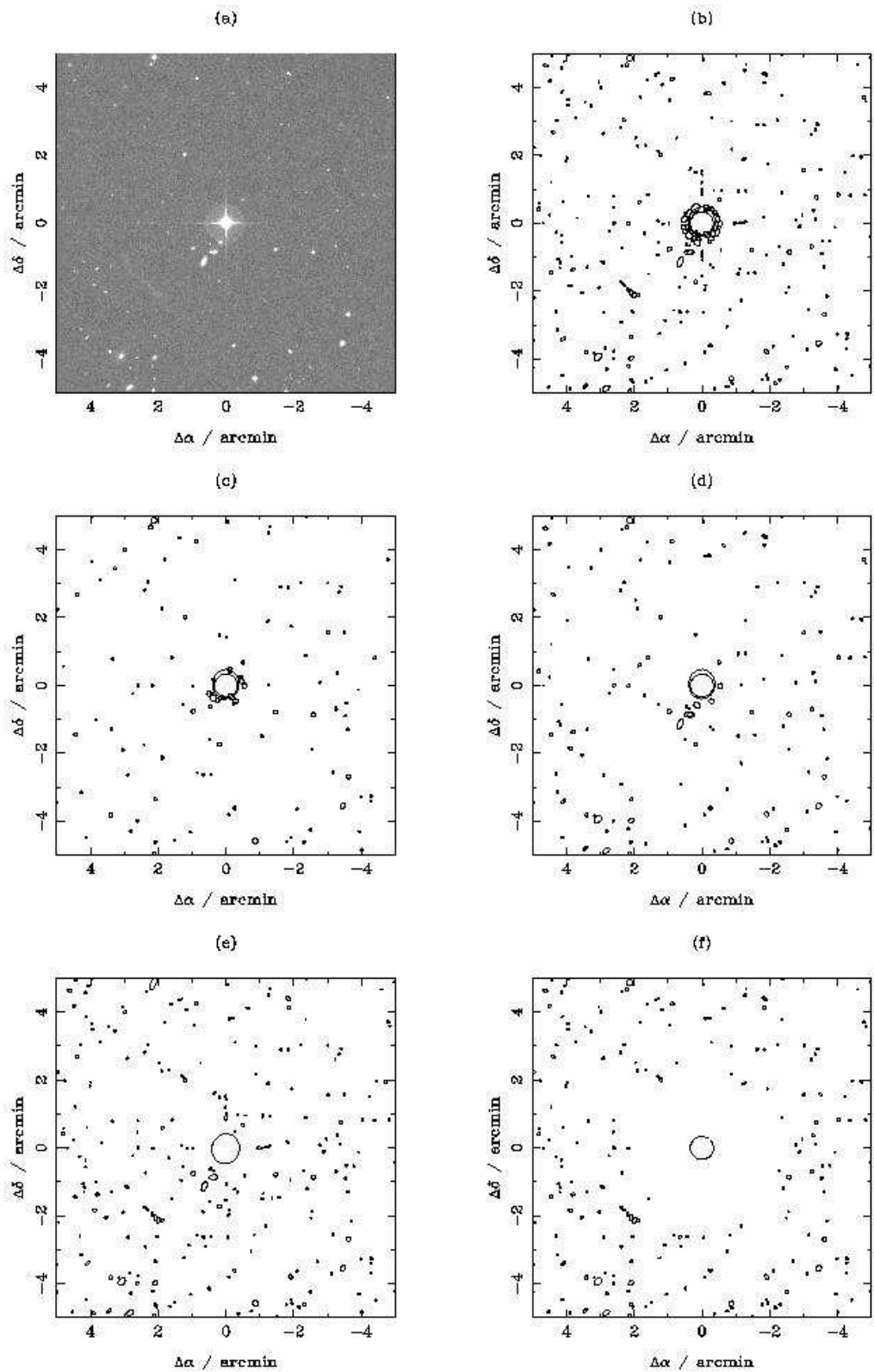}
\end{picture}
\caption[]{Spurious objects near a bright (m=10) star: (a) J--band pixel data;
(b) all images detected; (c) images left after an
ellipticity cut $e\leq0.25$; (d) images having an R--band pair; (e) images
left when deblending is not used; and (f) images left when using the quality
flag to exclude all detections in the vicinity of the bright star. In
(b), (c), (d) and (f) deblending is used (i.e.~child images of blends are
shown).}
\label{bright}
\end{figure*}

In Figure~\ref{bright}(a) and~(b) we show B$_{\rm J}$ image data in the 
vicinity of a bright (m~$\sim10$) star along with and ellipse plot of the 
raw image catalogue
from those data. The remaining panels of Figure~\ref{bright} show ellipse plots
of object catalogues created by (c) performing an ellipse `cut' of,
somewhat arbitrarily, 
$e\leq0.25$ (where $e=1-b/a$ and $a,b$ are the semi--major and semi--minor
ellipse axes); (d) by using the R--band magnitude to exclude stars that are
unpaired between the 
B$_{\rm J}$ and~R plates; (e) by switching off the deblending 
(i.e.\ including only those images with blend flag~$\leq0$); and (f) by using
the quality flag bit~10 to exclude all images within a circular 
area encompassing the
spurious images (Paper~{\sc II}, Section 2.2.2). Clearly, there are advantages
and disadvantages to each of these methods. For example, using the quality flag
to `drill' around bright stars is conservative in that {\em all} images 
(including real ones) are expunged from the circular region, while the
ellipticity cut necessarily removes real galaxy images that are elliptical.
The final method employed depends on the application, and must be left to the
user, since there is a compromise to be made between contamination by
spurious images and completeness of real objects.

\section{Conclusion}
\label{concs}

We have presented a description of the SuperCOSMOS Sky Survey (SSS). We have
described the properties of the SSS in relation to those of other
digitisation programmes and presented examples of the use of the data.
With reference to external data, we have illustrated the accuracy 
and reliability of SSS image parameters. Along with Papers~{\sc II} 
and~{\sc III} we have produced a comprehensive description and user guide 
for the SSS. Table~\ref{global} summarises the global properties of
SSS object catalogue data.
\begin{table*}
\begin{center}
\begin{tabular}{cccc}
Astrometric properties: & \multicolumn{2}{c}{Absolute} & \\
 & \multicolumn{2}{c}{accuracy} & units \\
\multicolumn{4}{l}{ }\\
$\alpha,\delta$ (B$_{\rm J}<19$) & \multicolumn{2}{c}{0.1} & arcsec \\
$\alpha,\delta$ (faint images) & \multicolumn{2}{c}{0.3} & arcsec \\
proper motion $\mu_{\alpha,\delta}$ (R~$<17$) & \multicolumn{2}{c}{10.0} & mas~yr$^{-1}$ \\
$\mu_{\alpha,\delta}$ (faint images) & \multicolumn{2}{c}{50.0} & mas~yr$^{-1}$ \\
zeropoint error $\mu_{\alpha,\delta}$ (R~$<17$) & \multicolumn{2}{c}{$<10.0$} &  mas~yr$^{-1}$\\
zeropoint error $\mu_{\alpha,\delta}$ (R~$>17$) & \multicolumn{2}{c}{$\leq1.0$} &  mas~yr$^{-1}$ \\
\multicolumn{4}{l}{ }\\
Photometric properties: & \multicolumn{2}{c}{Accuracy} & \\
 & absolute & relative & units \\
\multicolumn{4}{l}{ }\\
$\sigma_{\rm B,R,I}$ ($<19,18,17$) & 0.3 & 0.05 & mag \\
$\sigma_{\rm B,R,I}$ (faint images) & 0.3 & 0.3 & mag \\
$\sigma_{\rm (B-R)}$ (B$_{\rm J}<17$) & 0.07 & 0.07 & mag \\
$\sigma_{\rm (B-R)}$ (faint images) & 0.16 & 0.16 & mag \\
\multicolumn{4}{l}{ }\\
Image detection/ & external & external & \\
completeness     & completeness & reliability & \\
\multicolumn{4}{l}{ }\\
B$_{\rm J}<19.5$ & $\sim100$\% & $\sim100$\% & \\
B$_{\rm J}\sim21$, R~$\sim19$ & $\sim75$\% & $\sim90$\% & \\ 
\multicolumn{4}{l}{ }\\
\end{tabular}
\caption[ ]{Global properties of SSS object catalogue data (for these
purposes, `relative' means within restricted position and magnitude 
ranges). For more detailed information see 
Papers~{\sc II} and~{\sc III}.}
\label{global}
\end{center}
\end{table*}

At the time of writing, SSS sky coverage consists of full southern hemisphere
(i.e.\ $\sim20000$ square degrees) in B$_{\rm J}$ and~R; first--epoch R coverage
is presently limited to the $\sim5000$ square degree SGC region while I~band
coverage is somewhat less than this. Projected dates for full completion of the
southern hemisphere are $\sim$~mid 2002. At present, it is unclear whether
the scanning programme will be extended into the northern hemisphere.

Apart from possible extension to whole--sky coverage, our future plans are
as follows. Enhanced database access allowing fast, general queries to the
entire dataset are being investigated; sophisticated database organisation
(e.g.\ Szalay \& Brunner~1997) and object--oriented design (Brunner~1997)
are promising approaches. `Release' of stand--alone, seamless object
catalogues (e.g.\ for support operations at observatories) is desirable, and
can be done when the survey is complete to some specified level. Other enhancements currently underway include addition of short exposure plates
at low Galactic latitude (the UK Schmidt `short red' survey, Hartley \& 
Dawe~1981) for more reliable astrometry in crowded regions, and in future
may include photometrically calibrated images, and proper motions based on the
early epoch POSS--I `E' plates for $\delta\geq-20^{\circ}$. To a certain extent,
enhancements and extensions depend on user demand from the community, and 
interested parties are encouraged to contact us directly if requiring
large--scale exploitation of these data.
The SSS homepage is \verb+http://www-wfau.roe.ac.uk/sss+.

\section*{Acknowledgements}

The SuperCOSMOS Sky Survey project owes its existence and success to a large 
number of individuals. On the hardware side, many people at the former Royal
Observatory at Edinburgh (now the United Kingdom Astronomy Technology Centre)
were involved; particular thanks are due to Bill Cormack, Lance Miller,
Magnus Paterson, Jim Herd, Janette Jameson, Tom Paul, Richard Bennett and
Joel Sylvester. On the software side, we acknowledge the support of Steven
Beard, Clive Davenhall, Bernard McNally and Mike Irwin. Obviously, the surveys
would have been impossible without the diligence of many observers
at the Schmidt telescopes, and many talented photographers in the development
and copying laboratories. On the UK Schmidt side, we acknowledge the efforts of
Malc Hartley, Ken Russell, Russell Cannon, Debi Allan, Fenella 
Stuart--Hamilton and Jason Cowan.
Funding for the Wide Field Astronomy Unit at the Institute for
Astronomy at the University of Edinburgh is provided by the UK PPARC. We
acknowledge the use of {\sc Starlink} computing facilities at the Institute for
Astronomy. We are indebted to the referee, Sean Urban, for a prompt and 
thorough review of these manuscripts.

The National Geographic Society--Palomar Observatory Sky Survey (POSS--I) was
made by the California Institute of Technology with grants from the National
Geographic Society. The UK Schmidt Telescope was operated by the Royal
Observatory Edinburgh, with funding from the UK Science and Engineering 
Research Council (later the UK Particle Physics and Astronomy Research Council),
until 1988~June, and thereafter by the Anglo--Australian Observatory. The blue
plates of the southern sky atlas and its equatorial extension (together known
as the SERC--J/EJ) as well as the Equatorial Red (ER), the second epoch
(red) Survey (SES or AAO--R) and the infrared (SERC--I) Survey
were taken with the UK Schmidt Telescope. All data retrieved from the URLs
described herein
are subject to the copyright given in this copyright summary. Copyright
information specific to individual plates is provided in the downloaded FITS
headers.

\vfill
\bsp

\end{document}